\documentclass[english]{iopart}
\usepackage[T1]{fontenc}
\usepackage[latin9]{inputenc}
\usepackage{amstext}
\usepackage{amssymb}
\usepackage{graphicx}
\usepackage{esint}

\makeatletter
\usepackage{iopams}
\usepackage{setstack}

\makeatother

\usepackage{babel}
\begin{document}

\title[Octupole deformation in the actinides]{Octupole deformation properties of actinide isotopes within a mean
field approach}

\author{L M Robledo}

\address{Dep. Física Teórica, Módulo 15, Universidad Autónoma de Madrid, }

\address{E-28049 Madrid, Spain}

\ead{{\large luis.robledo@}uam.es}

\author{R. R. Rodríguez-Guzmán}

\address{Department of Chemistry, Department of Physics and Astronomy }

\address{Rice University, Houston, Texas 77005, USA}
\begin{abstract}
We discuss the octupole deformation properties of many even-even actinide
isotopes. The analysis is carried out in the mean field framework
(Hartree-Fock-Bogoliubov approximation) by using the axially symmetric
octupole moment as a constraint. A one-dimensional octupole collective
Hamiltonian is used to obtain properties like excitation energies
or $E1$ and $E3$ transition probabilities of the negative parity
band-heads associated to the lowest lying $1^{-}$ and $3^{-}$ states.
The evolution of these values with neutron number is discussed and
a comparison with available experimental data is made. In order to
minimize the uncertainties associated to the energy density functional
used, the calculations have been carried out for an assorted set ranging
from the BCP1 functional to the finite range Gogny interaction with
the D1S, D1N and D1M parametrization.
\end{abstract}



\ams{44}

\submitto{\JPG }

\maketitle

\section{Introduction}

Nowadays it is well established both theoretical and experimentally
that octupole deformation is responsible for some unusual properties
of the low energy spectrum of several isotopes of radium and thorium
as are the presence of low lying $1^{-}$ states in even-even nuclei
\cite{Butler.96}. As those nuclei are usually quadrupole deformed,
the next member of the negative parity rotational band, namely the
$3^{-}$ is also observed. This $3^{-}$ is connected by strong $B(E3,3^{-}\rightarrow0^{+})$
transition probabilities to the ground state. Another typical feature
of octupole deformation is the appearance of alternating parity rotational
bands. Although there are several examples of parity alternating rotational
bands at low spins \cite{Butler.96,Phillips.96} the parity alternation
usually appears at high spin as a consequence of the stabilizing effect
of angular momentum on the octupole character of the system \cite{Garrote.97,Garrote.98}
(moments of inertia increase with the octupole moment and therefore
configurations with higher octupole moments are favored as angular
momentum increases). 

The appearance of octupole effects is strongly linked to the underlying
single particle spectrum \cite{BohrMott.75}. The reason is that octupolarity
is enhanced when, in a given major shell, the intruder interacts (via
a particle-hole excitation) with a standard parity orbital with three
units less of angular momentum ($3\hbar$ is the amount of angular
momentum carried out by the octupole operator). This happens between
the $j_{15/2}$ and $g_{9/2}$, the $i_{13/2}$ and $f_{7/2}$ and
the $h_{11/2}$ and $d_{5/2}$ orbitals. For those regions where both
protons and neutrons feel a strong octupole interaction is where octupole
related effects are expected. For example, in the region around $_{88}^{224}$Ra
the proton's Fermi level is close to the $f_{7/2}$ orbital that interacts
strongly with the empty $i_{13/2}$ nearby. Besides, the neutron's
Fermi level is near the $g_{9/2}$ orbital that strongly interacts
through the octupole interaction with the $j_{15/2}$ nearby. This
combined tendency of both protons and neutrons towards octupole deformation
is responsible for the observed features of negative parity states
in the region around $^{224}$Ra (see \cite{Butler.96} for relevant 
bibliography). An interesting question is whether
the octupole deformation effects persists when we increase proton
number and move away from the ideal case of Ra. For neutrons we already
know the answer as octupole effects are only strong for the radium
isotopes with mass numbers in the limited range 222-228 (N=134-140).
However, neutrons are in a different major shell in the actinides
and the dependency with proton number is far from being the same.
To answer this question we have carried out a series of calculations
in several isotopes of the even-even actinides U, Pu, Cm, Cf, Fm and
No exploring the possibility of octupole deformed ground states. To
this end, we have computed potential energy curves (PECs) as a function
of the axially symmetric octupole moment (K=0 bands) within the constrained
Hartree-Fock-Bogoliubov (HFB) approximation \cite{Ring.80,Bender.03}.
The PECs together with the collective octupole inertia are used as
input for a one-dimensional collective Hamiltonian calculation that
allows the evaluation of the energy of the lowest lying $1^{-}$ and
$3^{-}$ states as well as $B(E1)$ and $B(E3)$ transition probabilities
to the $0^{+}$ground state \cite{Egido.89,Egido.90}. Several parametrizations
of the Gogny \cite{Decharge.80,Berger.84} and BCP \cite{Baldo.08,robledo.08,Robledo.10,baldo.10}
energy density functionals (EDF) are used. 
%--------------------------------------------------------------
The reason for the consideration of this variety
of interactions is its empirical and phenomenological 
nature. The interactions are fitted to bulk properties
of nuclei (binding energies, radii, etc) and not to 
the spectroscopic quantities considered in this paper.
Therefore, a variation of their values with the interaction 
considered is to be expected. The range of variation with
different interactions can be taken as an estimator
of the error attributable to the uncertainties in the nuclear interaction.

All the nuclei considered
in this work have a deep prolate minimum along the quadrupole degree of freedom.
As a consequence the coupling between the quadrupole and octupole degrees of 
freedom is expected to be weak and therefore the octupole degree
of freedom alone is going to play a role in the
properties of negative parity states \cite{Robledo.88}. There are other regions of the
periodic table \cite{Skalski.93,zberecki.06,Yamagami.01}
where the quadrupole-octupole coupling as well as triaxiallity effects are relevant.
%---------------------------------------------------------------------------

Recently a theoretical survey of K=0 octupole states for even-even
nuclei has been published \cite{Robledo.11b}. In this survey, the
Generator Coordinate Method (GCM) with the octupole degree of freedom
has been used to compute octupole properties of over 800 even-even
nuclei including some of the nuclei discussed in this paper. However,
a large fraction of the nuclei studied here are quite close to proton's
drip line and were not considered in \cite{Robledo.11b}.

\section{Theoretical tools}

The description of octupole properties will be carried out in the
mean field framework \cite{Ring.80} with the realistic Gogny \cite{Decharge.80,Berger.84}
and BCP \cite{Baldo.08,robledo.08,Robledo.10,baldo.10} EDFs. As pairing
correlations play a relevant role in the low energy nuclear dynamics
the full HFB approximation will be considered. In this scheme, quasiparticle
annihilation and creation operators are defined as linear combinations
of a conveniently chosen single particle basis. In our case, we have
used a Harmonic Oscillator (HO) basis which has been chosen big enough
as to warrant convergence of the physical quantities (energies, transition
probabilities, etc) with the basis size. To solve the constrained
HFB equations an approximate second order gradient method \cite{Robledo.11a},
based on the parametrization of the energy in terms of the Thouless
expansion of the most general HFB wave functions and using two quasiparticle
energies as precondititioner, has been used. Axial symmetry has been
preserved in the calculation suggesting the use of an axially symmetric
HO basis made up of the tensor product of two dimensional HO wave
functions times one-dimensional HO ones. Along with the octupole moment
constraint associated to the multipole operator $\hat{Q}_{3}=r^{3}Y_{30}$
and used to generate the potential energy curves (PEC's), we have
included a constraint on the center of mass of the nucleus (i.e. the
mean value of $r^{1}Y_{10}$ has been set to zero) to prevent the
coupling to the spurious center of mass motion. As a consequence of
the axial symmetry imposed in the HFB wave functions the remaining
components $\mu\ne0$ of the corresponding multipole operators $r^{3}Y_{3\mu}$
and $rY_{1\mu}$ have a zero mean value by construction. 

The information given by mean field theories is restricted to the
energy and shape of the -generally- deformed ground state. To restore
the parity symmetry broken by the solutions of the mean field approximation
and in order to describe the dynamics of the collective excited states
of octupole character it is mandatory to go beyond the mean field
approximation. With this in mind, the octupole degree of freedom $Q_{3}=\langle\psi|\hat{Q}_{3}|\psi\rangle$
(where $|\psi\rangle$ is the HFB intrinsic wave function) has been
used to build up a collective Hamiltonian based on the GCM and the
Gaussian Overlap Approximation (GOA) \cite{Brink.68,Giraud.74,Reinhard.87}.
In this method, the GOA is used to reduce the Hill-Wheeler equation
of the GCM to a Schrödinger equation for the collective wave function,
the so-called Collective Schrödinger Equation (CSE) 
\begin{equation}
\hat{{\cal \mathcal{H}}}_{coll}\phi_{\alpha}(Q_{3})=\epsilon_{\alpha}\phi_{\alpha}(Q_{3}),\label{SCHCOL}
\end{equation}
 where the collective Hamiltonian $\hat{{\cal H}}_{coll}$ is given
by 
\begin{eqnarray}
\hat{{\cal \mathcal{H}}}_{coll} & = & -\frac{1}{\sqrt{G(Q_{3})}}\frac{\partial}{\partial Q_{3}}\sqrt{G(Q_{3})}\frac{1}{2B(Q_{3})}\frac{\partial}{\partial Q_{3}}\label{HCOLL}\\
 & + & V(Q_{3})-\epsilon_{0}(Q_{3}).
\end{eqnarray}
 In this expression $G(Q_{3})$ is the metric, $B(Q_{3})$ is the
mass parameter associated with the collective motion along $Q_{3}$,
$V(Q_{3})$ is the collective potential given by the HFB energy $V(Q_{3})=\langle\psi(Q_{3})|\hat{H}|\psi(Q_{3})\rangle$
and $\epsilon_{0}(Q_{3})$ is the Zero Point Energy (ZPE) correction.
The eigenfunctions $\phi_{\alpha}(Q_{3})$ of equation (\ref{SCHCOL})
have to be normalized to one with the metric $G(Q_{3})$ 
\begin{equation}
\int dQ_{3}\,\sqrt{G(Q_{3})}\,\phi_{\alpha}^{*}(Q_{3})\phi_{\beta}(Q_{3})=\delta_{\alpha,\beta}
\end{equation}
 in order to preserve the hermiticity of $\hat{{\cal \mathcal{H}}}_{coll}$.

It has to be mentioned that a collective Schrödinger equation can
also be obtained from the Adiabatic Time Dependent Hartree-Fock (ATDHF)
theory \cite{Baranger.78,Brink.76,Villars.77} after quantization
of the semi-classical Hamiltonian for the slow moving collective degrees
of freedom. The collective Hamiltonian obtained in this way has the
same functional form as the GCM+GOA one, but the expression of the
collective parameters is different. Later on we will discuss how to
choose these collective parameters.

An interesting property of $\hat{{\cal H}}_{coll}$ is its invariance
under the exchange $Q_{3}\rightarrow-Q_{3}$ that allows the classification
of its eigenfunctions, $\phi_{\alpha}(Q_{3})$, according to their
parity under the $Q_{3}\rightarrow-Q_{3}$ exchange. It is easy to
see that the parity of the collective wave function under the $Q_{3}\rightarrow-Q_{3}$
exchange corresponds to the spatial parity operation in the correlated
wave function built up from $\phi_{\alpha}$. The inclusion of octupole
correlations immediately restores the parity symmetry lost at the
mean field level \cite{Egido.89,Egido.91}. Therefore, the solution
of the CSE equation (\ref{SCHCOL}) allows the calculation of the
$0^{+}-1^{-}(3^{-})$ energy splitting as well as the $B(E1)$ and
$B(E3)$ transition probabilities connecting them. At this point it
has to be pointed out that in the present framework where only time
reversal invariant wave functions are considered it is only possible
to describe excited states with average angular momentum zero. To
deal with genuine $1^{-}$ or $3^{-}$ states a projection onto good
angular momentum should be performed, which is out of the scope of
the present work. Here we will assume that the cranking rotational
energy of the $1^{-}$state is much smaller than the excitation energy
of the negative parity band head and therefore can be safely disregarded.
With this approximation in mind, the reduced transition probabilities
from the lowest $1^{-}$ and $3^{-}$ states to the $0^{+}$ ground
state can be computed within the Rotational Model approximation as
\begin{equation}
B(E\lambda,I_{f}\rightarrow I_{i})=e^{2}\langle I_{i}K\lambda0|I_{f}K\rangle^{2}|\langle\varphi_{i}|r^{\lambda}Y_{\lambda,0}|\varphi_{f}\rangle|^{2},
\end{equation}
 where $|\varphi_{i}\rangle$ and $|\varphi_{f}\rangle$ are correlated
wave functions obtained in the spirit of the GCM from the collective
wave functions $\phi_{\alpha}(Q_{3})$. The above formula can be reduced
to an expression involving those collective wave functions $\phi_{\alpha}(Q_{3})$
by means of the GOA \cite{Nerlo.87}. The final result for $K=0$
bands reads 
\begin{equation}
B(E1,1^{-}\rightarrow0^{+})=\frac{e^{2}}{4\pi}\left|\langle\phi_{0^{-}}|D_{0}|\phi_{0^{+}}\rangle_{\textrm{COLL}}\right|^{2}\label{BE1}
\end{equation}
 for the E1 electric transition and 
\begin{equation}
B(E3,3^{-}\rightarrow0^{+})=\frac{e^{2}}{4\pi}\left|\langle\phi_{0^{-}}|Q_{30}(\textrm{PROT)}|\phi_{0^{+}}\rangle_{\textrm{COLL}}\right|^{2},\label{BE3}
\end{equation}
 for the E3 one. In the above formulas we have introduced the collective
matrix element of an operator $\hat{O}$ as
\[
\langle\phi_{0^{-}}|\hat{O}|\phi_{0^{+}}\rangle_{\textrm{COLL}}=\int dQ_{3}\, G^{1/2}\,\phi_{0^{-}}^{*}(Q_{3})O(Q_{3})\phi_{0^{+}}(Q_{3})
\]
 where $O(Q_{3})=\langle\psi(Q_{3})|\hat{O}|\psi(Q_{3})\rangle$.
In the formula (\ref{BE1}) $D_{0}$ is the dipole moment operator
whose mean value is defined as the difference between the center of
mass of protons and neutrons 
\begin{equation}
D_{0}(Q_{3})=\frac{N}{A}\langle\psi(Q_{3})|\hat{z}_{prot}|\psi(Q_{3})\rangle-\frac{Z}{A}\langle\psi(Q_{3})|\hat{z}_{neut}|\psi(Q_{3})\rangle.\label{eq:dipole}
\end{equation}
Finally, $Q_{30}(\textrm{PROT)}$ is the part of the octupole operator
acting on proton's space. In the recent survey of \cite{Robledo.11b}
it was pointed out the inadequacy of the rotational formula for the
calculation of E1 and E3 transition probabilities for near spherical
nuclei (see \cite{egido.96} for a beautiful example of how angular
momentum projection solves a related problem near $^{208}$Pb ). This
is not a problem in the present calculation as the nuclei considered
are well deformed and the rotational formula works well there.

To carry out the collective calculations it is necessary to specify
the collective parameters $G(Q_{3})$, $B(Q_{3})$ and $\epsilon_{0}(Q_{3})$
appearing in the definition of $\hat{\mathcal{H}}_{coll}$ equation
(\ref{HCOLL}). As it was mentioned before, there are two sets of
parameters coming from the GCM+GOA and the ATDHF derivation of the
collective Hamiltonian. The set of parameters used in this calculation
is an admixture of the two and it is known as the ATDHF+ZPE set. It
includes the mass parameter $B(Q_{3})$ coming out from the semi-classical
Hamiltonian of the ATDHF theory, the metric of the GCM+GOA and the
ZPE correction computed with the GCM+GOA formula but using the ATDHF
mass instead, i.e. 
\begin{equation}
\epsilon_{0}(Q_{3})=\frac{1}{2}G(Q_{3})B(Q_{3})_{ATDHF}^{-1}.\label{ZPE}
\end{equation}
 This set of parameters was devised to put together the advantages
of the ATDHF set (time-odd components included in the mass term) and
the ones of the GCM+GOA (ZPE correction). This method can be somewhat
justified in the context of the extended Generator Coordinate method
\cite{Reinhard.87,Villars.75} and has been extensively used \cite{Berger.84,Egido.89}.
In mean field calculations it is customary to include in the ZPE correction
the rotational energy. We have tested that the inclusion of the rotational
energy correction (as computed in \cite{egido.04} to take into account
effects near sphericity and the approximate calculation of the Yoccoz
moment of inertia) does not modify substantially the results, as could
be anticipated due to the weak coupling of the octupole and quadrupole
degrees of freedom.

The calculation of the collective parameters involves the inversion
of the HFB stability matrix which is closely related to the matrix
of the RPA equation. At present, this is a formidable task and approximations
are needed. The approximation used in this paper -- called {}``cranking
approximation\textquotedbl{} \cite{Reinhard.78,Girod.79} -- neglects
the off-diagonal terms of the stability matrix allowing to invert
it analytically but at the cost of including the two body interaction
only through the mean field. Although this approximation has been
extensively used in the literature for the calculation of collective
masses and moments of inertia (see for instance \cite{Baran.81,Berger.84,Boning.85})
its validity has not been properly established. Using the {}``cranking
approximation'', the ATDHF+ZPE parameters are given by 
\begin{equation}
G(Q_{3})=\frac{M_{-2}(Q_{3})}{2M_{-1}^{2}(Q_{3})},\,\,\, B(Q_{3})=\frac{M_{-3}(Q_{3})}{M_{-1}^{2}(Q_{3})};\label{eq:COLLM}
\end{equation}
 where the quantities $M_{-n}(Q_{3})$ ($n=1,2,3$ ) are defined as
\begin{equation}
M_{-n}(Q_{3})=\sum_{k,l}\frac{\left|\left(Q_{30}\right)_{kl}^{20}\right|^{2}}{(E_{k}+E_{l})^{n}}.\label{eq:Mn}
\end{equation}
 In the above expression, $E_{k}$ are the quasi-particle energies
and $\left(Q_{30}\right)_{kl}^{20}$ are the matrix elements of the
$20$ part \cite{Ring.80} of the octupole operator $\hat{Q}_{30}$
in the quasi-particle basis of the HFB wave function $|\psi(Q_{3})\rangle$.
This form of the collective mass is usually referred to in the literature
as the Belyaev-Inglis inertia \cite{Ring.80}.

\subsection{Interactions}

In this paper we have used the three modern parametrizations of the
Gogny functional \cite{Decharge.80} which are available in the market,
namely Gogny D1S \cite{Berger.84}, D1N \cite{Chappert.08} and D1M
\cite{Goriely.09}. There are mainly two reasons to proceed in this
way. First, to gain confidence on the independence of our results
concerning tiny details of the interaction/parametrization used. In
this respect, we have to comment that although the three Gogny's parametrizations
used belong to the same functional form of the interaction, their
different behavior regarding nuclear matter, pairing properties, binding
energies or radii suggest their non-equivalence. The other reason
to do the calculations with the D1N and D1M parametrizations is to
check their performance in describing octupole deformation phenomena
as a partial test to asses their ability to describe low energy nuclear
phenomena (see for instance \cite{RRRG.10a,RRRG.10b} for recent studies
with D1M). In addition to the Gogny force, we have used the recently
proposed BCP \cite{Baldo.08} energy density functional with the BCP1
parameter set \cite{Baldo.08}. This functional is based on a local
density approximation to a realistic nuclear matter equation of state
supplemented by a finite range surface term. The spin-orbit and Coulomb
interactions have the standard form as used in the Gogny force. The
pairing force used is a zero range, density dependent force that was
fitted to reproduce the pairing gap in nuclear matter given by the
Gogny functional. The parameters have been fitted to reproduce binding
energies and radii of selected spherical nuclei and the results show
a reasonable level of agreement comparable or even better than the
results provided by D1S. Several spectroscopic quantities have been
computed in several regions of the periodic table and the results
are in close agreement with the experimental values and the theoretical
results of D1S \cite{robledo.08,baldo.10}. Octupole properties of
Ra, Th and Ba isotopes have been analyzed in the past with this functional
\cite{Robledo.10} and the good agreement obtained at a qualitative
level in the comparison with the experiment and Gogny force calculations
is a clear indication of the suitability of this EDF for the exploration
of the octupole degree of freedom in the heavy actinides.

\section{Results}

\begin{figure}
\includegraphics[width=0.489\textwidth]{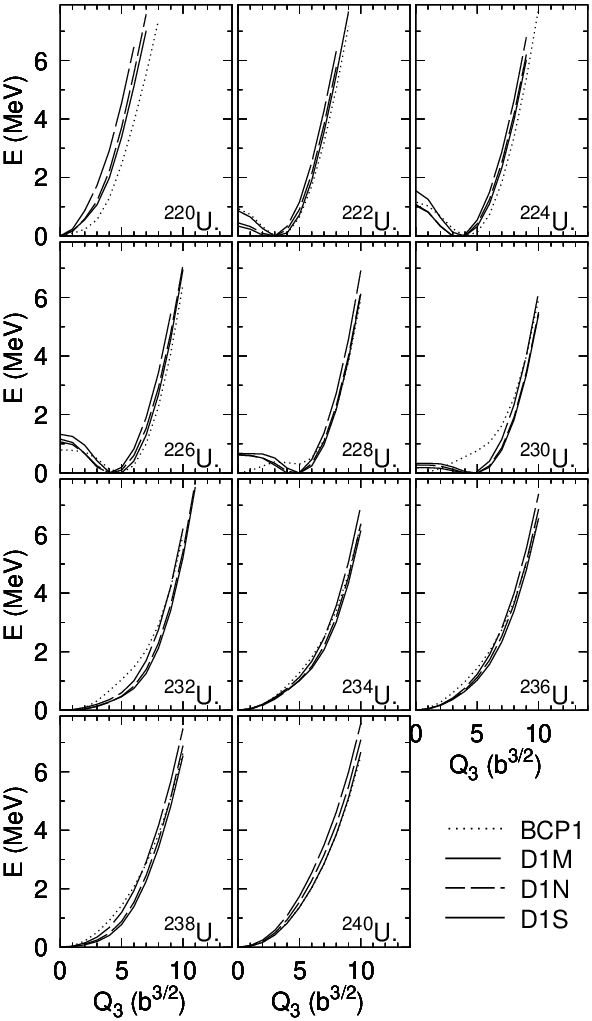}\includegraphics[width=0.5\textwidth]{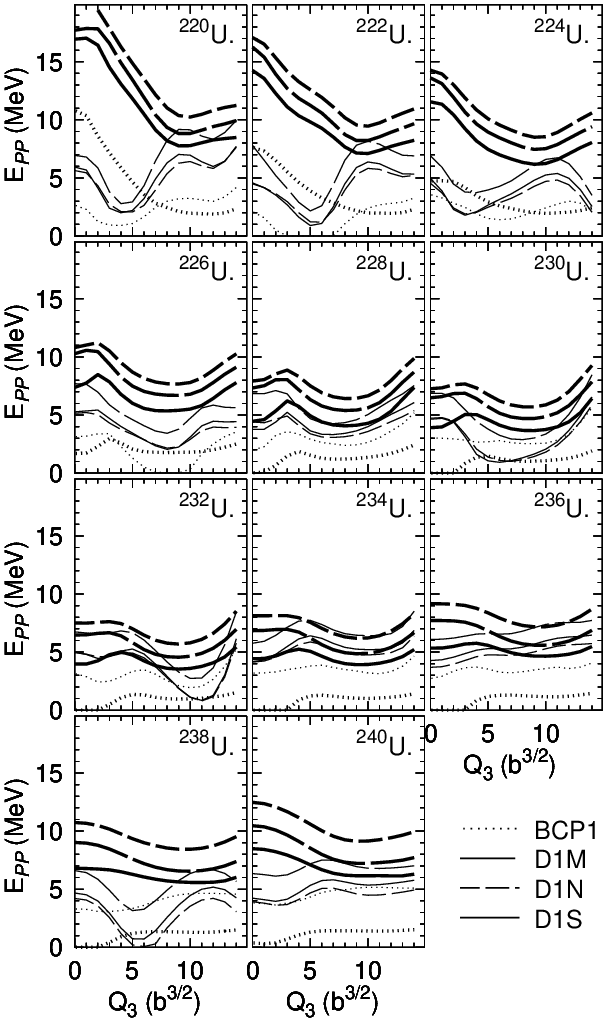}\caption{On the left hand side panels, the PEC's as a function of the octupole
moment $Q_{3}$ (in thousands of Fermi cube or $b^{3/2}$) for the
even-even isotopes of uranium with $A$ in the range between $A=220$
and $A=240$. Results for the D1S (full lines), D1N (dashed lines),
D1M (dashed dotted lines) parametrizations of the Gogny and the BCP1
(dotted lines) functionals are presented. On the right hand side panels
the particle-particle correlations energies for both protons (thick
curves) and neutrons (thin curves) are given as a function of the
octupole moment in $b^{3/2}=10^{3}$fm$^{3}$.\label{fig:MFPES_U}}
\end{figure}

In the eight isotopic chains considered we have followed the same
procedure, first the octupole constrained HFB wave functions have
been generated along the lines described in the previous section.
Next the collective mass and zero point energies are computed with
the mean field wave functions and together with the PEC itself they
are used to build the Collective Schrödinger Hamiltonian. From the
lowest lying solutions of the CSE we obtain the excitation energy
of the first negative parity excited state (a $1^{-}$ for these deformed
nuclei) and the $B(E1,1^{-}\longrightarrow0^{+})$ and $B(E3,3^{-}\longrightarrow0^{+})$
transition probabilities (the $3^{-}$ state in the $B(E3)$ transition
probability refers to the second member of the rotational band built
on top of the $1^{-}$band head). The whole procedure will be thoroughly
described for the case of the uranium isotope. For the other isotopic
chains only the most relevant results, namely the PES and the observables
obtained after solving the CSE will be presented.

\subsection{Low excitation energy properties of the uranium isotopes}

In fiugre \ref{fig:MFPES_U} we display, as a function of the octupole
moment $Q_{3}$ and the four interactions used, the PECs (left hand
side panels) and the particle-particle correlation energies (right
hand side panels). The particle-particle correlation energy is defined
as $E_{pp}=\frac{1}{2}\textrm{Tr}\Delta\kappa$, and it is given in
terms of the pairing field $\Delta$ and the pairing tensor $\kappa$.
This quantity gives a rough idea of the amount of pairing correlations
in the system. It can also be used as an indicator of where the single
particle level density around the Fermi surface is high or low as
high level densities are associated to strong pairing correlations.
This quantity is also correlated with the pairing gap that represents
the energy of the lowest two quasiparticle excitation and therefore
it is closely related to the collective inertias to be discussed below.
Coming back to the figure, the PECs obtained with the four interactions
considered are very similar. For the particle-particle correlation
energies the same similarity is obtained for the three parametrization
of Gogny, but for the BCP functional the particle-particle correlation
energies are significantly smaller. Of the eleven isotopes considered,
five ($^{222-230}$U) show an octupole deformed minimum with a depth
never exceeding the 1.5 MeV. The octupole moment of the minima changes
with the neutron number in a non-trivial way. The remaining nuclei
with a minimum at $Q_{3}=0$ show a parabolic behavior centered around
the minimum with a curvature that is more or less independent of the
interaction considered. There seems to be a certain correlation between
the appearance of octupole deformed minima and the minima of the neutron
pairing energies of the right hand side panels. This is something
to be expected as the presence of deformed minima is usually related
(Jahn-Teller effect) to regions of low level density of the underlying
single particle orbitals and these low level density regions usually
correspond to small pairing correlations.

\begin{figure}
\includegraphics[width=0.5\textwidth]{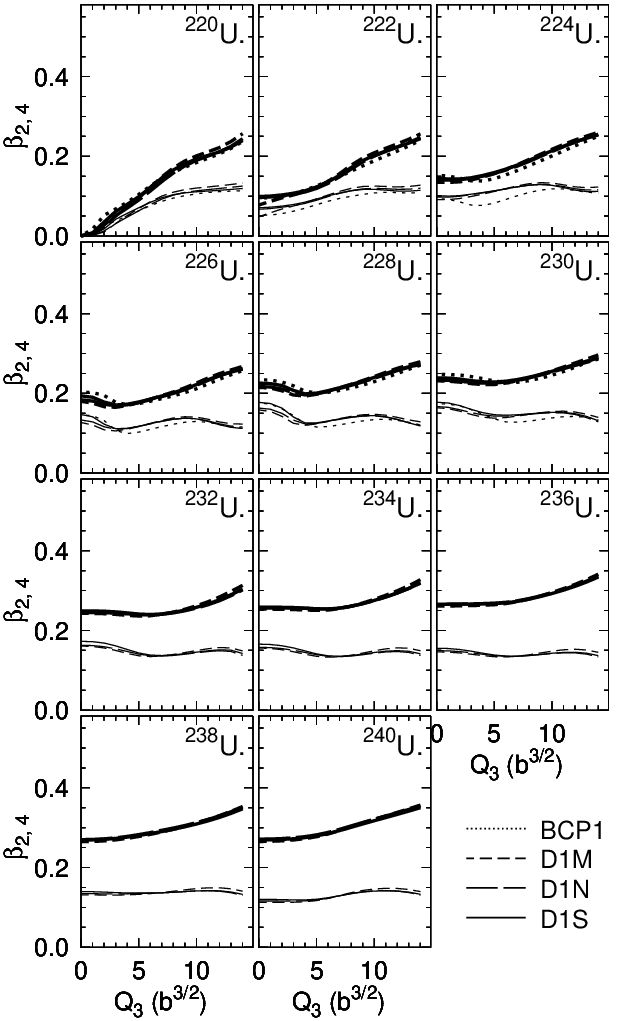}\includegraphics[width=0.5\textwidth]{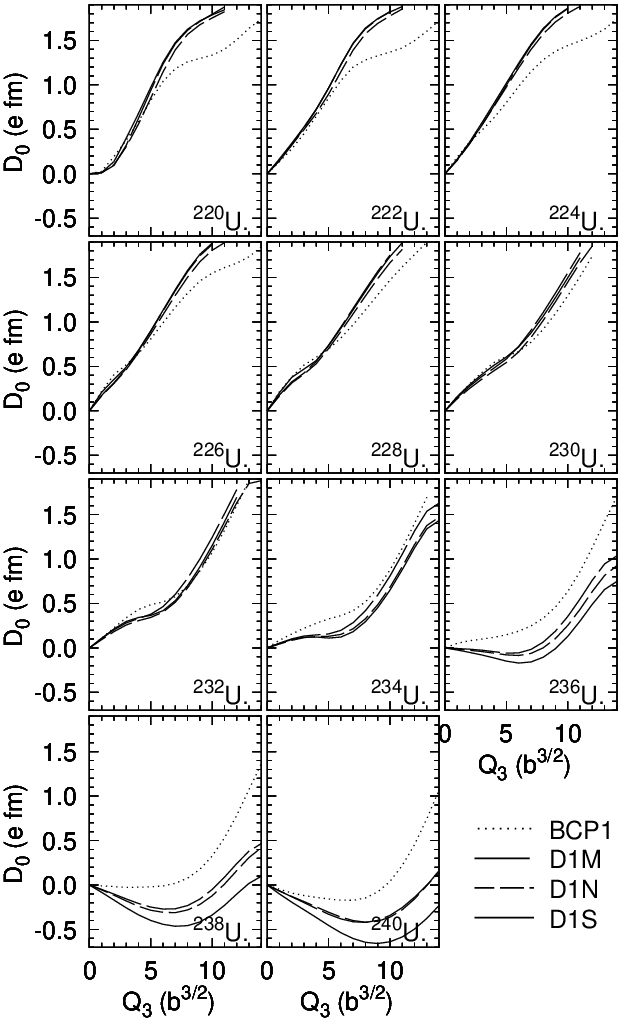}\caption{On the left hand side panels, the $\beta_{2}$ (thick lines) and $\beta_{4}$
(thin lines) deformation parameters are plotted as a function of the
octupole moment $Q_{3}$ (in $b^{3/2}\equiv10^{3}$fm$^{3}$) for
the even-even isotopes of uranium with $A$ in the range between $A=220$
and $A=240$. Results are given for the D1S (full lines), D1N (dashed
lines), D1M (dashed dotted lines) and BCP1 (dotted lines) functionals.
On the right hand side panels the dipole moment is shown as a function
of $Q_{3}$.\label{fig:MFMULT_U}}
\end{figure}

As the octupole moment increases, the nuclear shape changes and the
values of other multipole moments including the quadrupole, hexadecapole
and dipole moments change. In the left hand side panels of figure
\ref{fig:MFMULT_U} the $\beta_{2}$ and $\beta_{4}$ deformation
parameters are represented as a function of $Q_{3}$. For most of
the U isotopes considered the variation with $Q_{3}$ is small and
it is only in the lightest nuclei $^{220}$U and $^{222}$U where
the changes are noticeable. The nucleus $^{220}$U is spherical at
$Q_{3}=0$ and develops a sizable quadrupole deformation with increasing
octupole moment. On the left hand side panels of this figure the dipole
moment of equation (\ref{eq:dipole}) is plotted. The dipole moment
is zero by construction at $Q_{3}=0$ and its behavior with the octupole
moment is determined by the occupancies of specific single particle
orbitals around the Fermi level. High $K$ orbitals give the largest
contributions to $D_{0}$ and therefore, depending on their occupancy
for protons or neutrons as a function of $Q_{3}$, we can observe
different behaviors. For the lightest isotopes the dipole moment increases
almost linearly with $Q_{3}$. The slope decreases with neutron number
to the point that for $^{236}$U the value of $D_{0}$ is almost zero
for a wide range of octupole moments as a consequence of competing
neutron and protons contributions. Given the intimate connection between
the dipole moment and the $B(E1)$ transition probabilities we conclude
that the situation found in $^{236}$U will lead to a quenched $B(E1)$
value as compared to the ones of other isotopes of the same nucleus. 

\begin{figure}
\includegraphics[width=0.5\textwidth]{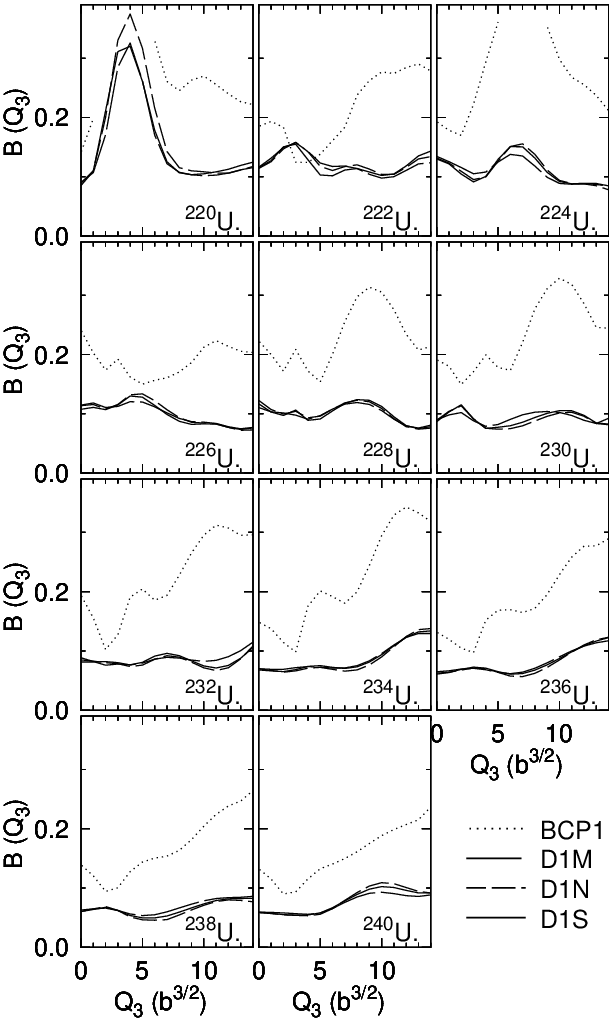}\includegraphics[width=0.5\textwidth]{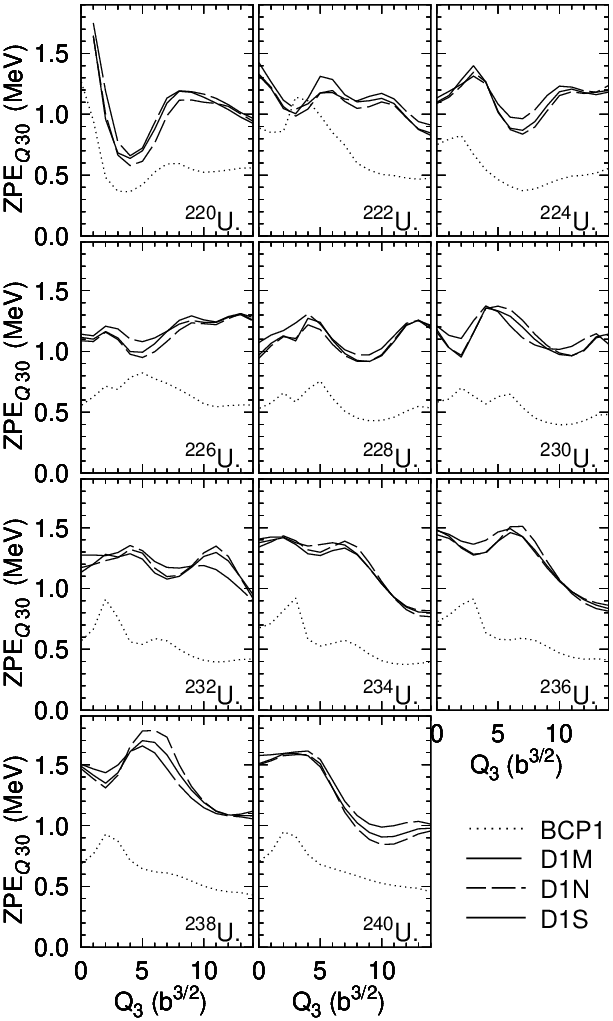}\caption{On the left hand side panels the octupole collective inertia parameter
$B(Q_{3})$, which is one of the relevant ingredients in the one-dimensional
collective Schrödinger Hamiltonian (see text for details), is shown
as a function of $Q_{3}$ for the considered U isotopes. On the right
hand side panels the zero point energy (ZPE) correction of equation
(\ref{ZPE}) is represented as a function of $Q_{3}$.\label{fig:CollM_U}}
\end{figure}

In figure \ref{fig:CollM_U} the collective inertia $B(Q_{3})$ of
equation (\ref{eq:COLLM}) is depicted as a function of $Q_{3}$.
It plays a central role in the outcome of the collective Schrödinger
Hamiltonian calculations mentioned in the previous section. The collective
inertia is roughly speaking inversely proportional to the pairing
correlations (the pairing gap to be more specific) and therefore the
similarity between the collective masses obtained for D1S, D1N and
D1M indicate that a subtle cancellation between proton's and neutron's
contributions has to take place (see the particle-particle correlation
energies in figure \ref{fig:MFPES_U}). It is also worth to notice
that the peaks observed in the $B(Q_{3})$ plots are related to regions
of low pairing correlations. On the other hand, the BCP1 inertia is
systematically higher than the ones of the different parametrizations
of the Gogny functional. The difference amounts to factors of two
or even more in some specific cases. This will be responsible for
the differences observed in the solution of the CSE.

\begin{figure}
\includegraphics[scale=0.5]{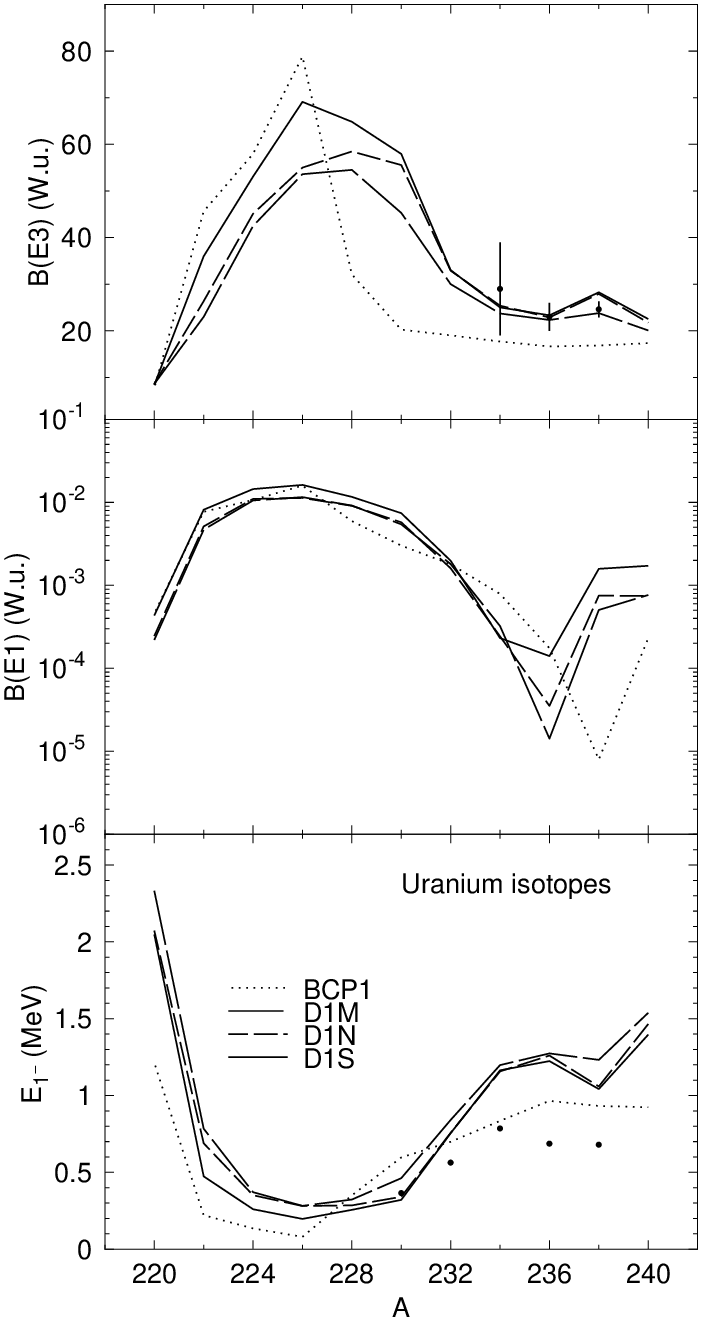}\includegraphics[scale=0.5]{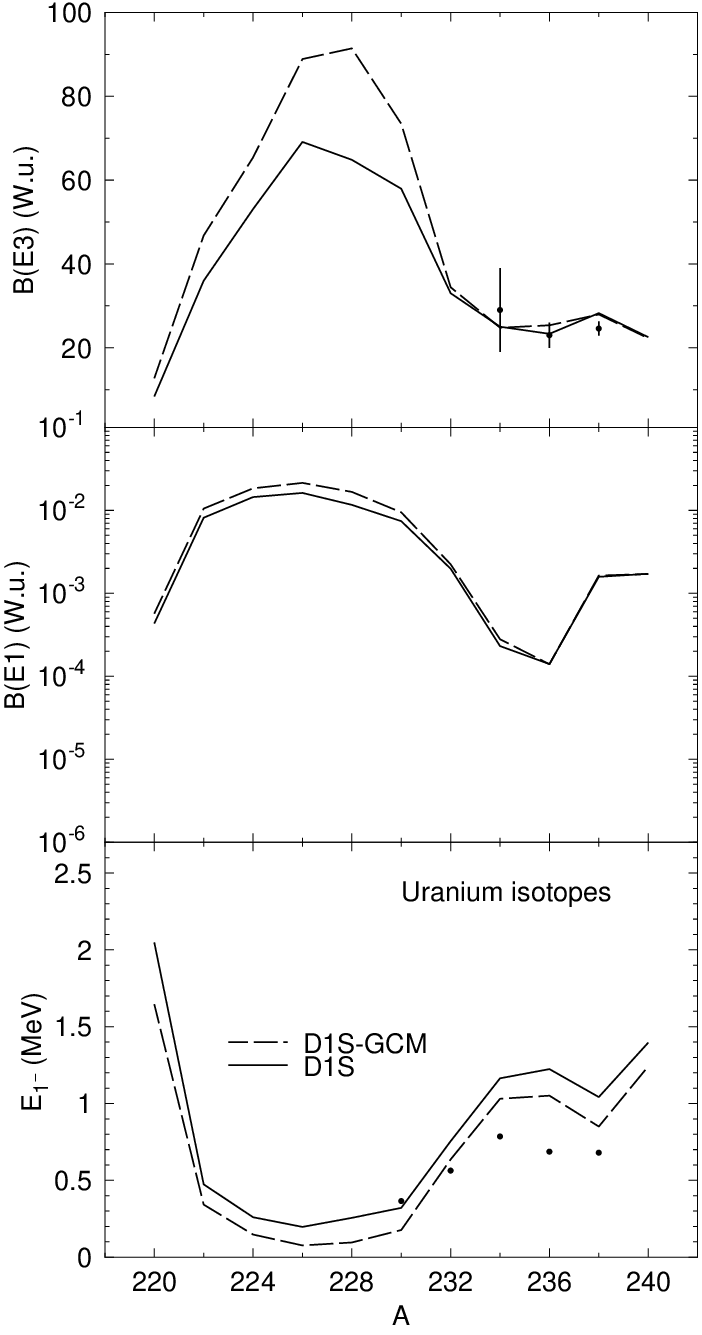}

\caption{Left hand side panels: Results of the 1D collective Schrödinger Hamiltonian
calculations for the three parametrizations of Gogny (D1S, D1N and
D1M) as well as for the BCP1 energy density functional. In the lowest
panel the energies of the $1^{-}$ states as a function of the mass
number for the uranium isotopes are depicted. Theoretical results
(lines) are plotted along with experimental data (dots). In the middle
panel the $B(E1,1^{-}\rightarrow0^{+})$ transition probabilities
in W.u. are given. Finally, in the upper panel the $B(E3,3^{-}\rightarrow0)$
are also given, for the different isotopes and different theoretical
prediction, in W.u. In the right hand side panels a comparison is
made between the results obtained in the present approach and the
ones obtained with Gogny D1S in the framework of the GCM for the octupole
degree of freedom.\label{fig:RES_U}}
\end{figure}

On the right hand side panels of figure \ref{fig:CollM_U} the zero
point energy correction $\epsilon(Q_{3})$ of equation (\ref{ZPE})
is depicted for the isotopes of uranium. The $\epsilon(Q_{3})$ values
are correlated with the inverse of the collective inertia $B(Q_{3})$
as can easily be noticed in the figure and its behavior with $Q_{3}$
can modify somehow the topology of the PES but not at the level of
changing significantly the physical results. On the average the $\epsilon(Q_{3})$
values are of the order of 1.5 MeV which is consistent with the correlation
energies for parity symmetry restoration and octupole dynamics obtained
in \cite{Robledo.11b}. This energy together with the energy gain
obtained by the breaking of reflection symmetry at the mean field
level (of the order of one MeV at most) should be added to the nucleus
binding energy but the relatively constant values as a function of
mass number indicate that the impact of such addition in separation
energies is not expected to be significant. After solving the one-dimensional
collective Schrödinger equation we obtain the excitation energies
of the $1^{-}$ states and the $B(E1,1^{-}\rightarrow0^{+})$ and
$B(E3,3^{-}\rightarrow0^{+})$ transition probabilities. The results
obtained with Gogny D1S, D1N and D1M as well as for the BCP1 energy
density functional for the U isotopes are depicted along with available
experimental values \cite{Kibedi.02}in the left hand side panels
of figure \ref{fig:RES_U}. For all the results considered we observe
that the three parametrizations of the Gogny EDF produce the same
isotopic trend being the differences not very relevant from a physical
point of view. Obviously, the differences are larger for the most
sensitive quantities, namely the transition probabilities. The $1^{-}$
excitation energies are very small for the $^{222}$U to $^{230}$U
isotopes. This is the expected behavior as those nuclei show permanent
octupole deformation at the HFB level (see figure \ref{fig:MFPES_U}).
For the other U isotopes, the HFB energy shows a minimum at $Q_{3}=0$
with a parabolic behavior characteristic of octupole vibrational states
which lie higher in energy, as observed in the energies of the negative
parity excited states of figure \ref{fig:RES_U}. The comparison with
experimental data is reasonable as we are able to reproduce the increasing
octupolarity as neutron number decreases. Concerning the $B(E1,1^{-}\rightarrow0^{+})$
transition probabilities we observe a pronounced minimum around $A=236$
that is a direct consequence of the behavior of the dipole moment
with $Q_{3}$ observed in figure \ref{fig:MFMULT_U}. Unfortunately,
there is no experimental data available for these isotopes concerning
the $B(E1)$ transition probabilities and the associated dipole moments
\cite{Butler.91}. The $B(E3,3^{-}\rightarrow0^{+})$ transition probabilities
show a marked maximum around $A=226$ that is correlated to the minimum
in the energies of the $1^{-}$ states. For the heavier isotopes the
computed $B(E3)$ compare well with experimental data. On the other
hand, the BCP1 results compare reasonably well with the Gogny ones
except for the $^{228-232}$U isotopes where the octupole correlations
are weaker and therefore the $1^{-}$ excitation energies are higher
and the $B(E3)$ transition probabilities lower. Also the minimum
observed in the $B(E1)$ as a function of neutron number is shifted
two units of neutron number when compared with the Gogny results.
In the right hand side panels, a comparison is made between the theoretical
results obtained in the present framework and the ones of the GCM
with the octupole degree of freedom as generating coordinate (GCM-Q3)
obtained for Gogny D1S in \cite{Robledo.11b}. In both calculations
the underlying mean field is the same and therefore it should not
be surprising the agreement between the results of both calculations.
The differences observed in the excitation energies and $B(E3)$ transition
probabilities can be attributed to the different collective masses
used in each of the calculations %
\footnote{In the GCM-Q3 calculation there is no explicit collective mass as
the full Hill-Wheeler equation is solved but, if a local approximation
to the non-local HW equation is considered the concept of collective
mass can be recovered, see Ref \cite{Ring.80} for a discussion.%
}. However, the magnitude of the differences is compatible with the
magnitude of the differences attributed to the use of different parametrizations
and/or functionals in the calculations. The message that the comparison
of all these theoretical predictions is telling us is that the behavior
of the physical quantities as a function of mass number is independent
of the kind of calculation and interaction and therefore can be taken
as a strong prediction. On the other hand, the values obtained for
the physical quantities oscillate typically in a range of plus or
minus 40 \% and therefore their values should be taken just as an
indication of the order of magnitude to be expected.

\subsection{Plutonium isotopes}

\begin{figure}
\includegraphics[width=0.52\textwidth]{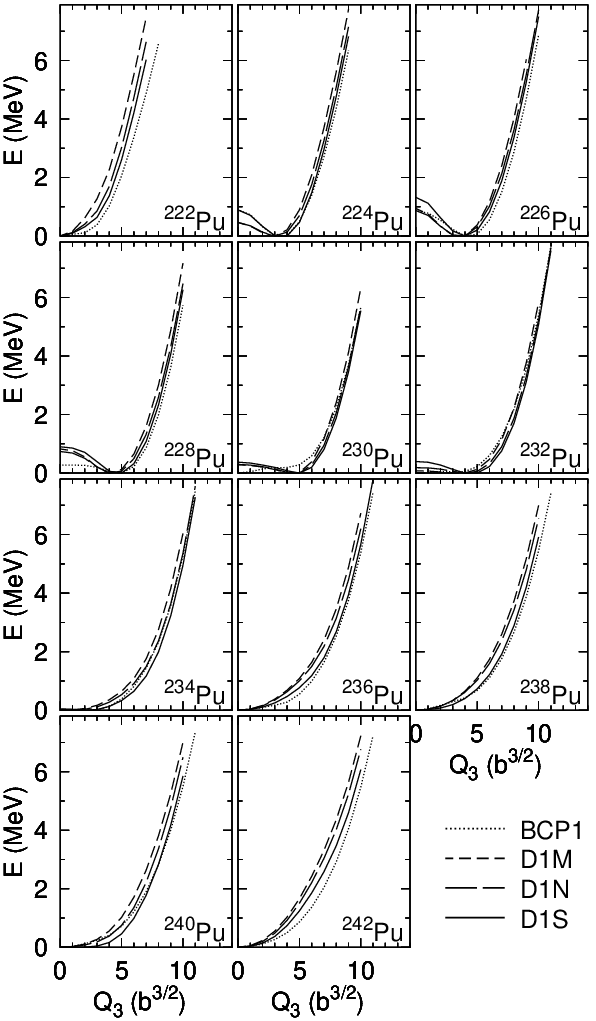}\includegraphics[width=0.48\textwidth]{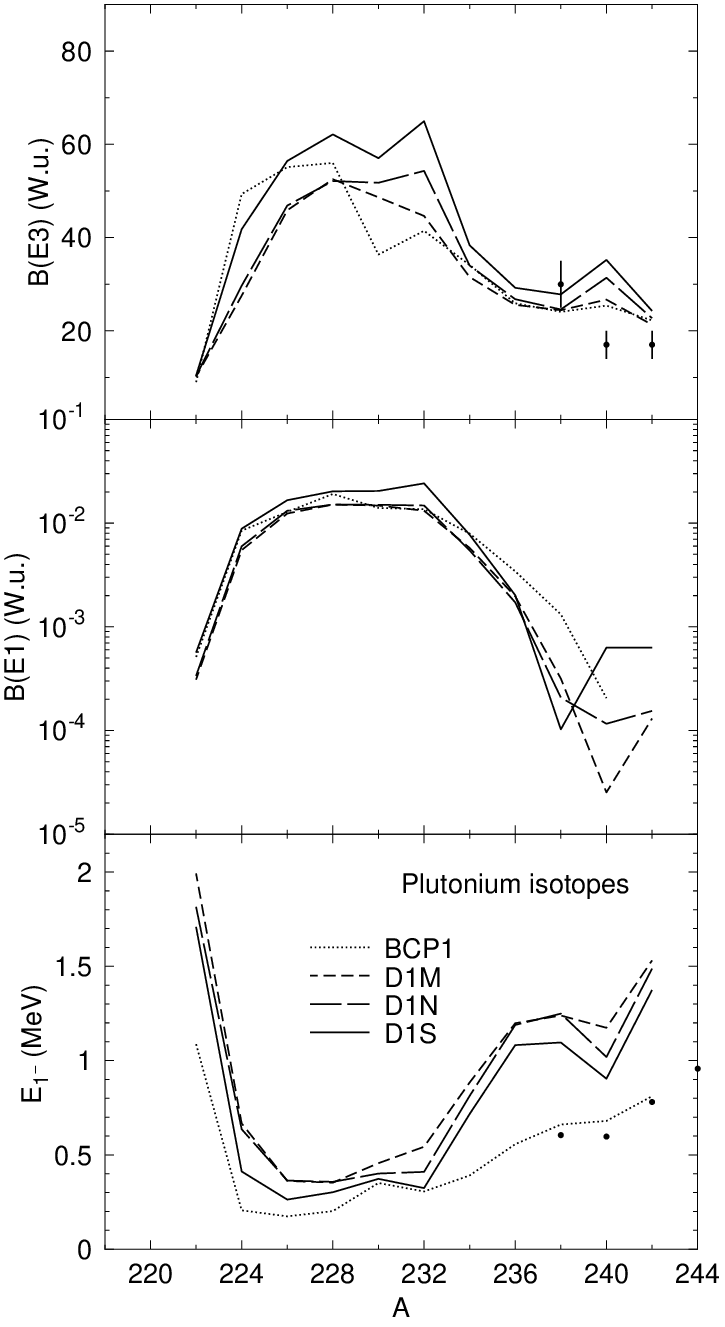}\caption{On the left hand side panels, the PES, as a function of $Q_{3}$,
are shown for the considered Pu isotopes. In the right hand side panels,
the $1^{-}$ excitation energies (in MeV) as well as the $B(E1)$
and $B(E3)$ transition probabilities (in W.u.) are plotted as a function
of the mass number $A$. Theoretical results are depicted as lines
of various types (see legends) and the experimental data, taken from
\cite{Kibedi.02}, are represented by bullets.\label{fig:MFPES_Pu}}
\end{figure}

Results for the Plutonium (Z=94) isotopes in the mass range 222-242
(corresponding to neutron numbers 128 -- 148 ) are discussed in this
section. For this and the remaining isotopic chains to be discussed,
we will focus on physical quantities like excitation energies and
transition probabilities and the only {}``intrinsic'' quantity to
be shown is the PES as a function of the octupole moment to quantify
the gain in binding energy associated to octupole correlations. In
figure \ref{fig:MFPES_Pu} we have plotted on the left hand side the
PES as a function of the octupole moment for the Pu isotopes considered.
Well deformed octupole minima are obtained for the isotopes $^{224}$Pu,
$^{226}$Pu and $^{228}$Pu with mean field octupole correlation energies
as large as 1 MeV in $^{226}$Pu. The nuclei $^{230-234}$Pu show
very shallow minima which are located at $Q_{30}=0$ in the heaviest
of the three. For the other isotopes, the minima are reflection symmetric
($Q_{30}=0$) showing a parabolic behavior as a function of $Q_{30}$
with a rather large curvature. On the right hand side the $1^{-}$
excitation energy as well as the $B(E1)$ and $B(E3)$ transition
probabilities (in Weisskopf units) are plotted as a function of mass
number. We observe that a range of isotopes starting with $^{224}$Pu
and ending with $^{232}$Pu show octupole deformation in their ground
state. The mean field octupole correlation energy is of the order
of 1 MeV in the lighter isotopes and gets reduced to a few hundred
keV for $^{232}$Pu. As a consequence, the $1^{-}$ excitation energy
for those five isotopes is of the order of 0.5 MeV and much smaller
than in the neighboring isotopes. As a result of the enhanced octupolarity
in these isotopes the $B(E3)$ transition probabilities are stronger
reaching values of up to 60 W.u. The $B(E1)$ transition probabilities
show a rather constant behavior for the octupole deformed isotopes
and show a dip for $^{238}$Pu in the case of D1S and $^{240}$Pu
in the case of both D1N and D1M.

\subsection{Curium results}

\begin{figure}
\includegraphics[width=0.52\textwidth]{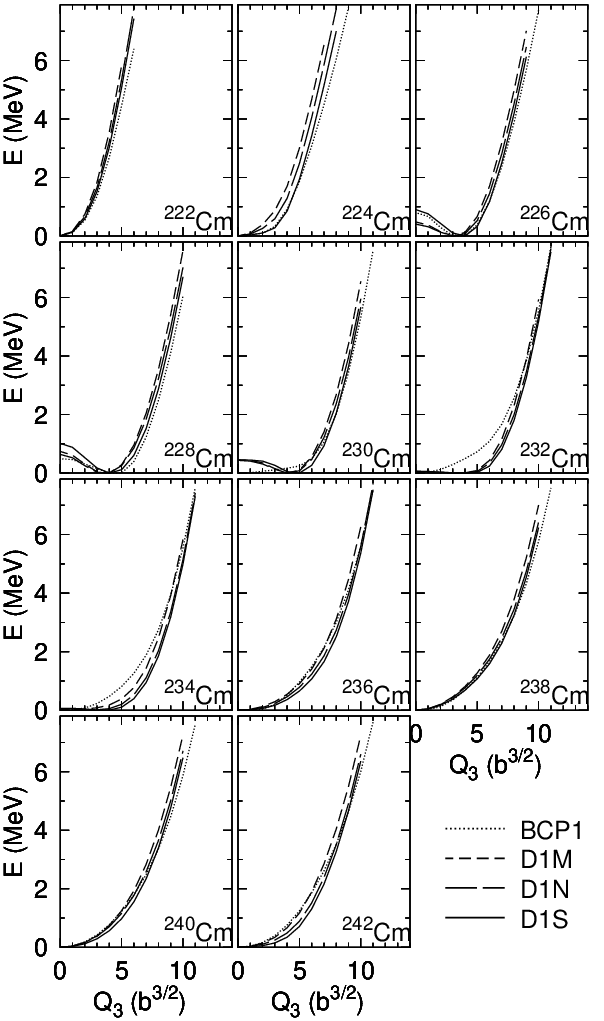}\includegraphics[width=0.53\textwidth]{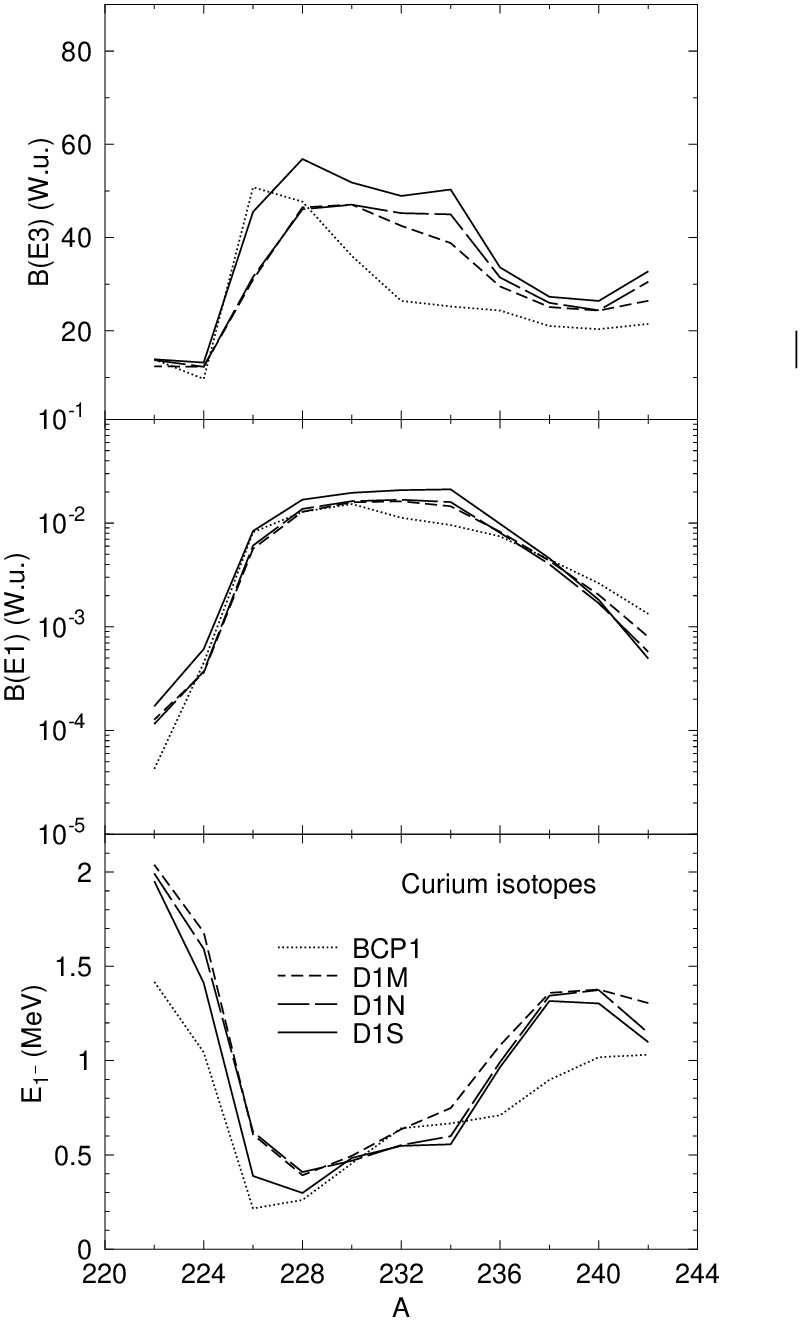}\caption{Same as figure \ref{fig:MFPES_Pu} but for the curium isotopes \label{fig:MFPES-Cm}}
\end{figure}

Results for the Curium (Z=96) isotopes with neutron numbers in between
N=126 and N=146 are discussed in this section. In figure \ref{fig:MFPES-Cm}
we have plotted on the left hand side the PES as a function of the
octupole moment. Octupole deformed minima are obtained for the nuclei
$^{226-230}$Cm with energy gains of the order of a few hundreds of
keV being $^{226}$Cm the nucleus with the deepest octupole deformed
well. For $^{232-234}$Cm the minima are at $Q_{30}=0$ but the minima
are in the two cases rather shallow. For the other nuclei the minima
is at $Q_{30}=0$ and the PES shows a parabolic behavior with a rather
large curvature. On the right hand side panels, the $1^{-}$ excitation
energy as well as the $B(E1)$ and $B(E3)$ transition probabilities
(in Weisskopf units) obtained after solving the 1D CSE are plotted
as a function of mass number. The results resemble at a qualitative
level the ones for the Pu isotopes indicating the relatively low impact
on octupole correlations of the addition of two protons. Strong octupole
correlations, with low lying $1^{-}$ states are anticipated to show
up in the mass range 226-232. In that range the $B(E3)$ transition
probabilities are rather collective and are around 50 W.u. For the
nucleus $^{234}$Cm the results show a dependency with the interaction,
pointing to a characterization of this nucleus as transitional in
terms of octupole correlations. 

\subsection{Californium}

\begin{figure}
\includegraphics[width=0.52\textwidth]{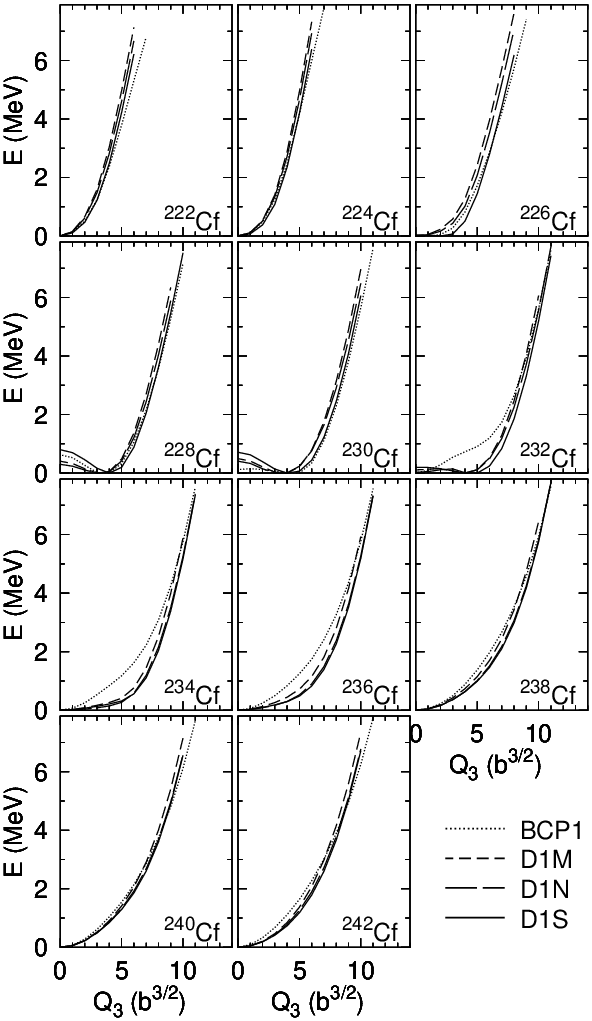}\includegraphics[width=0.61\textwidth]{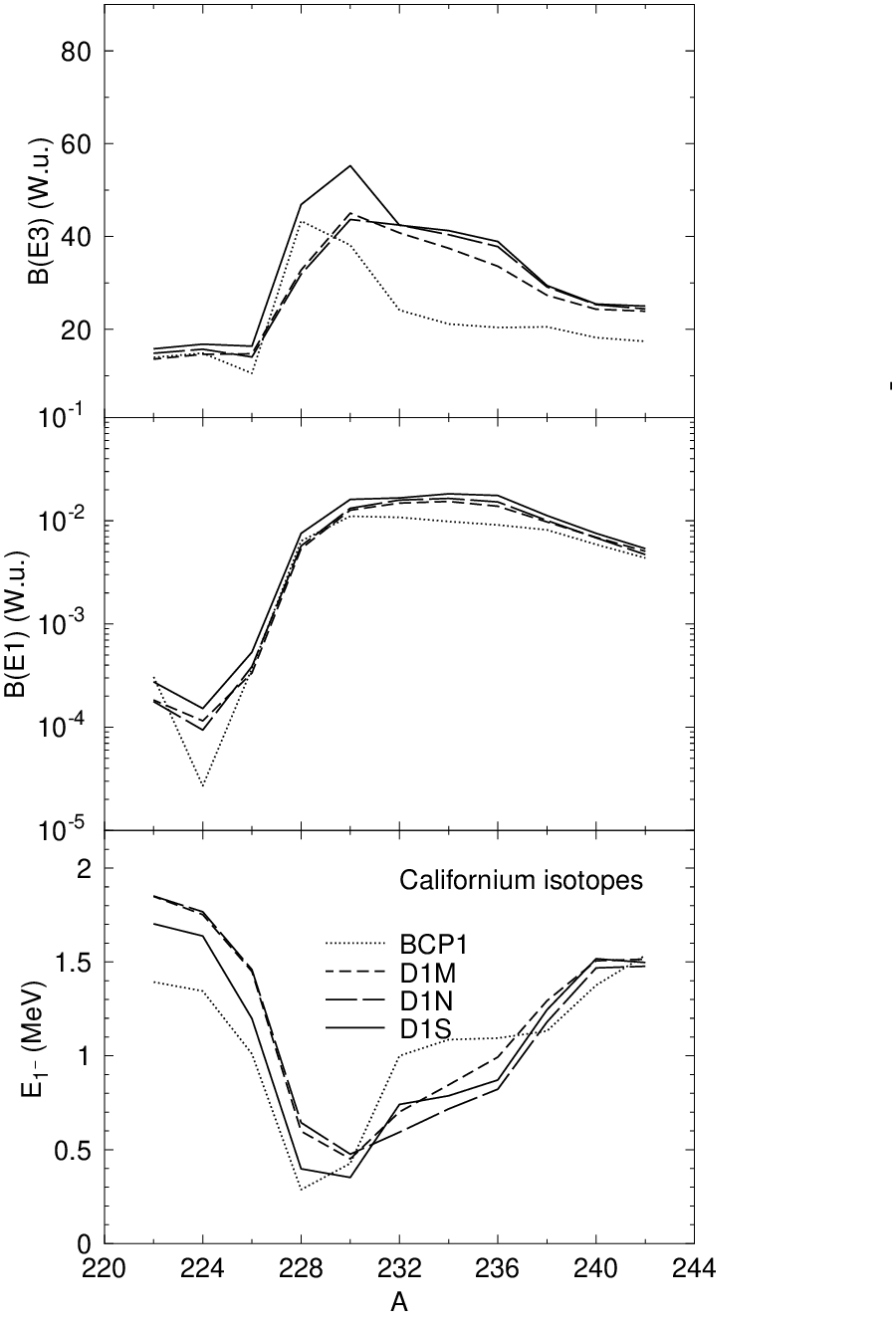}\caption{Same as figure \ref{fig:MFPES_Pu} but for the californium isotopes\label{fig:MFPES-Cf}}
\end{figure}

Results for the Californium (Z=98) isotopes with neutron numbers in
the 124 --144 range are presented and discussed in this section. In
figure \ref{fig:MFPES-Cf} we have plotted on the left hand side the
PES as a function of the octupole moment. On the right hand side the
$1^{-}$ excitation energy as well as the $B(E1)$ and $B(E3)$ transition
probabilities (in Weisskopf units) are plotted as a function of mass
number. The qualitative and even quantitative similarity with the
Cm results presented in the previous section is striking and indicates
that the addition of the two protons has little influence in the octupole
deformation properties. Strong octupole correlation effects are predicted
in the mass number range 228 - 236 and a severe quenching of $B(E1)$
transition probability is anticipated for the nuclei around $^{224}$Cf.
At this point it has to be mentioned that proton's drip line is predicted
to be at $^{228}$Cf in calculations with Gogny D1S \cite{Hilaire.07}.
Therefore it is rather unlikely that the octupole effects expected
in the range 228-236 will ever be accessible experimentally. 

\subsection{Fermium and Nobelium}

For the Fermium (Z=100) and Nobelium (Z=102) isotopes calculations
with neutron numbers in the range N=122 and N=142 have been carried
out. Although octupole deformation is present in our calculations
for the ground state of $^{230-232}$Fm ($N=130-132)$ and $^{232-234}$No
($N=130-132)$ we have decided not to present the results here as
those nuclei are right at the proton's drip line, according to calculations
with the Gogny D1S force (see \cite{Hilaire.07} for details), and
therefore it is rather unlikely that they will ever be created in
the laboratory.

\section{Conclusions}

We have performed systematic mean field calculations for a relevant
set of even-even actinides to explore the impact of octupole correlations
in the low energy nuclear spectrum. We observe strong octupole correlations
in some uranium, plutonium, curium and californium isotopes with a
strength that decreases with proton number. The neutron number of
the isotopes with strong octupole correlations become closer to the
proton's drip line as proton number increases. As a consequence, the
fermium and nobelium isotopes showing up noticeable octupole correlations
are beyond the drip line. Excitation energies for the $1^{-}$state
as low as 300 keV and $B(E3)$ transition probabilities of 60 to 70
W.u. are predicted in some uranium isotopes, making them good candidates
for experimental studies aiming to extend the region of known nuclei
where octupole correlation effects are important. On the theoretical
side, the similarity among the results obtained with three different
parametrizations of the Gogny functional give credit to them as suited
for an equivalent description of octupole correlation effects. The
similitudes between the Gogny and BCP results indicate that the octupole
correlation effects obtained are genuine and not an artifact of the
phenomenological interactions used.

\ack{}{}

This work was supported by MICINN (Spain) under research grants FIS2008-01301,
FPA2009-08958, and FIS2009-07277, as well as by Consolider-Ingenio
2010 Programs CPAN CSD2007-00042 and MULTIDARK CSD2009-00064.

\end{document}